# Chapter 1[1]

# Scientific Culture and Its Role in International Negotiations

*Klaus Gottstein*

The general hypothesis of this book claims that when representatives of a particular professional culture interact with representatives of another professional culture, their performance is conditioned by their respective common frame of reference. In other words, the course of such interaction will vary according to the profession of an envoy or delegate—be he a professional diplomat, a politician experienced in domestic affairs or, say, a scientist. In this chapter, this theory is applied to the field of science, examining specifically scientific culture and its role in international negotiations.

History offers many interesting examples of scientists acting as diplomatic negotiators. The great philosopher, mathematician, engineer, and geologist, Gottfried Wilhelm Leibniz (1646–1716), served as envoy to his prince elector, the Archbishop of Mainz, to persuade King Louis XIV of France to attack Egypt rather than the German Holy Roman Empire. He later traveled all over Europe, negotiating with Emperor Leopold I in Vienna and Czar Peter the Great, and was instrumental, through his historical research and personal connections, in securing for the House of Hanover first the electorship of the Holy Roman Empire and later the succession to the English throne. Another example is Benjamin Franklin (1706-1790), a famous scientist and diplomat. Not only did he research electricity and invent the lightning rod, he was also the U.S. envoy to England and France, and negotiated the treaty by which France agreed to support the American revolutionaries.

In these early examples, scientists were seemingly appointed to become negotiators of their respective countries because of their status as men of high standing. Later on, their special competence in certain fields seemed to be more relevant. For example, in 1940, a National Defense Research Committee was created in the United States. One year later it was merged with the Office of Scientific Research and Development (OSRD) within the Executive Office of the President. The director of the OSRD, Dr. Vannevar Bush (1890-1974), a renowned electrical engineer who developed the first electronic analogue computer, had direct access to the then U.S. president, Franklin D. Roosevelt, and enjoyed considerable influence within the government bureaucracy. A branch of the OSRD was also set up in London, while the United Kingdom opened the British Central Scientific Office (BCSO) in Washington. Later, Scientific Liaison Offices of Australia, New Zealand, Canada, and South Africa were established to form, with BCSO, the British Commonwealth Scientific Office. Its task was the coordination of cooperation with the U.S. in developing the atomic bomb, radar, and other systems (e.g., for warfare against submarines) in which modern science and technology played a leading part. The first directors of this office were internationally known scientists: Sir John Cockcroft, Sir Charles Darwin, Sir Thomas Taylor. Later, the head of the office received the title and responsibility of "Attaché for Scientific Questions" to the British Embassy in Washington. This institution was retained after the end of World War II.

---

[1] This chapter is part of the book „Professional Cultures in International Negotiation. Bridge or Rift?" (Editor: Gunnar Sjöstedt), Lexington Books, Lanham, MD, USA 2003, Copublished with IIASA, Copyright by International Institute for Applied Systems Analysis (IIASA). Thanks are due to Lexington Books, IIASA, and Ashgate Publishing Ltd for permission to reproduce here this text.

Events in the United States after World War II continued to give significant roles to scientists as international negotiators. In the late 1940s and in the 1950s, the U.S. government began sending university professors as science attachés to several of its European embassies. These academics were, however, almost exclusively representatives of American science, with only limited contact to the policymakers of their country.

The "Sputnik Shock" of 1957 in the United States changed this situation. The U.S. State Department set up a large science and technology division, with subdivisions for atomic energy, space technology, environmental protection, and general science policy. In addition, a new science attaché program was started. By 1970 the United States had science attachés in twenty-three countries. Other countries followed its example. In Washington alone, shortly thereafter, about twenty-five embassies had a science attaché on their staff. A typical agenda for a science attaché included observing developments, and negotiating agreements on cooperation in

- arms control and disarmament (but not weapons development, which came under the jurisdiction of the military attaché);

- energy and material resources;

- industrial productivity and competitiveness;

- information on scientific developments in their home countries for interested circles in the United States;

- education in science and technology;

- scientific solutions to social problems (environmental protection, public transport, urban development, reactor safety, health care, crime prevention, data processing, etc.); and

- science and technology policy.

## An Interconnected World

States, businesses, and other kinds of organization are confronted, in our modern world, with an international environment characterized by complex interdependence. Linear, monocausal thinking is no longer adequate under these conditions. Important events that occur in one part of the globe are immediately known around the world, and the repercussions of such disturbances also travel long distances. Television, computer networks, and air travel have shrunk distances dramatically. The emotions created by modern media with their reports on natural and political catastrophes in (formerly) distant lands, and the arrival on our doorstep of the refugees from such catastrophes, have deeply influenced political life in modern democracies. In front of a television set in the living room, the ordinary citizen can be "present" on the spot when world events happen. This has changed the roles of diplomats. While stationed in foreign countries and representing their governments in bilateral or multilateral negotiations, diplomats are in constant and immediate contact with their superiors at home for guidance. Thanks to modern means of communication and transportation, and due to interfaces between education, economy, ecology, social conditions, foreign relations, and national and international security, no single field of knowledge is sufficient for assessing the complicated by- and after-effects that any policy measure might have somewhere in the network. In many instances, interdisciplinary and international cooperation will be required for a satisfactory solution of the problems at hand. This situation is likely to increase the demand for scientific competence in negotiation and other forms of international cooperation.

Issues subject to negotiation have become increasingly science-related: arms control related to, nuclear, chemical and biological weapons, environmental protection, food and agriculture, water supply, health, energy, and population policy all have important scientific components. Consequently, the participation of scientific experts in international negotiations on these issues has become essential. Even in classical fields of diplomacy, such as those of conflict prevention and crisis management, there are now academic experts in political science and international relations or psychology who specialize in peace and conflict research. The knowledge required for a successful settlement, in international negotiations, of the political problems of our times is often of an interdisciplinary nature.

## Scientists as Bearers of Professional Culture in International Negotiations

Due to their education and training, as well as their professional accomplishments, working methods, and standards, scientists have their own professional culture, as explained in the Introduction chapter. The principal hypothesis of this book is that this professional culture conditions how individual scientists may generally color or have an impact on a particular negotiation. This impact may be produced in various ways. A key issue of this analysis is, hence, how scientists perform as negotiators in comparison to negotiators of other professional cultures, such as diplomats, military experts, or legal advisors.

According to the author's own practical experiences of complex international negotiations, the general advantages shared by scientists, as a group, include the following:

- Scientists have studied science, which represents a large portion of the knowledge of humankind.

- Scientists are used to dealing with complicated issues, to sorting out and analyzing facts, drawing conclusions, and revising them if new facts make revisions necessary. In many cases, they are able to propose alternatives for political decisions, estimating the costs, risks, and benefits of each option.

- Scientists can afford to think in terms of decades, to tackle long-term issues because, usually, they do not depend on re-election. An important task of scientists is to inform the public of the long-term consequences of specific policies.

- Scientists are familiar with international cooperation, since many scientific problems can only be solved in this manner. Cooperation creates ideas through the exchange of views, it facilitates the comparison of results, it stimulates criticism, it allows the formation of joint ventures and the realization of projects that would be beyond the scope of any one group or nation. It contributes toward an equilibrium of standards and forces, thereby reducing psychological and political tensions. International cooperation in science also opens up new channels of communication between different political systems.

These general characteristics of the scientist–negotiator may be interpreted as rough indicators of an international, scientific culture as compared to other professional cultures, for example that of the politician. Politicians in particular find it difficult to take on the long-term perspective that scientists are able to argue for. Politicians have to tackle the problems of today, they have to find the transition to tomorrow in a manner that is similar to how the "man or woman on the street" tends to handle his or her everyday problems. Politicians must win the next election: losing it would also mean losing their ability to influence the course of events.

Looking at the complex realities of international negotiations, it is not obvious how an intercultural confrontation between scientists and other professional cultures—such as that of politicians—should be solved most effectively. When scientists in one camp interact with scientists in another camp, their common background may clearly facilitate a negotiation. However, scientists do not automatically facilitate international negotiations in which they take part, at least not in a manner that politicians may appreciate. Scientists often do not have sufficient access to the realities of political life so that their advice sometimes disregards human and political factors when they put forward technical solutions to a political problem on the negotiation table.

Scientists are, of course, human beings like the rest of us. Usually, they love their work, are convinced of its importance, and want to achieve results. This may lead to a certain bias towards their own work. To counteract this inclination as far as possible, the institution of peer review has been introduced. The validity of scientific results, the importance of a scientific paper are not to be judged by the author himself but by some knowledgeable, usually anonymous, experts in the same field before a paper is accepted for publication, a project is funded, a prize is given, a professorship is offered, etc. In general, this works quite well.

An education in one of the natural sciences introduces a particular way of addressing and solving problems, indeed a certain mentality. Those who have been trained as scientists have learned to respect facts and the laws of nature that govern the physical world. They know that wishful thinking does not help to obtain results, but that conscientious work is required. Scientists know from experience that machines will stop running sooner or later unless they are supplied with energy and are carefully maintained. They have found out that there is a difference between mere assumption and firm knowledge, and they are familiar with the concept of probability. They can distinguish between predictable and unpredictable courses of events, and between developments that are likely to occur and those occurrences that just cannot be excluded. They respect proofs

5and are willing to change opinions that are proven wrong. They are convinced that deception does not pay because, sooner rather than later, the cheating is usually detected by those who try to confirm the feigned result. Most scientists are familiar with international cooperation and have colleagues in other countries with whom they have exchanged results or collaborated on joint projects. This has made them relatively immune to national prejudices as compared with some other professional cultures.

How do the properties of his, or her, professional culture condition the performance of a scientist who is asked to act as a negotiator? This author was in a position to make pertinent observations in this regard when he had the task of reassuring the U.S. State Department and the U.S. Atomic Energy Commission (AEC) of Germany's continued reliable partnership as a buyer of slightly enriched uranium for German power plants when, in the early 1970s, Germany for the first time decided to buy a certain amount of this fuel from the Soviet Union, where it was available at a lower price. The AEC, Germany's sole supplier till then, was upset, as was the U.S. State Department. These institutions were wondering whether, under the influence of the *Ostpolitik* of Chancellor Willy Brandt, the Federal Republic of Germany was slowly drifting away from the Western camp. It was possible to reassure them that this purchase was an entirely commercial, market-oriented step, and that German power plants would be happy to continue buying U.S. uranium, particularly when the price came down. In this connection a scientific argumentation based on facts and analysis had an important role to play.

Before joining the German Embassy in Washington, D.C., the author, in his capacity as a division leader in an independent research institute, had been accustomed to making his own decisions, after consultation with his colleagues, about which experiments should be carried out, which equipment should be bought, and similar professional matters. Sometimes negotiations had to be made with firms, or with the staff of the large accelerator centers about machine time at the accelerators where the equipment was to be exposed. Usually, there was collaboration with groups in other countries, which required agreement on the division of the data obtained, the standards of measurement, etc. This work was relatively unproblematic due to the important bridge-building effect that materializes when representatives from different countries but from the same professional culture meet at the negotiation table. A certain amount of bargaining was occasionally necessary, but, in general, agreement was easily reached. All parties shared the same goal: a successful scientific experiment yielding interesting results.

As a science attaché at the German Embassy in Washington, D.C., the author had to learn some new rules (Gottstein 1973, 435–444). First of all, when he wrote a report about his work, the information he had received, or the results of his negotiations, it was the ambassador, not he, who signed the report. The ambassador might even change some of the wording into more diplomatic language. And the report was always sent to the Foreign Ministry in Bonn, with a copy to the Ministry of Research and Technology, and not to anybody else who might also have been interested. Who else would be informed was decided by Bonn.

After a while the author got used to these and some other rules that seemed to contradict the standards of his own professional culture. He understood that in order to be effective as a negotiator and/or a government advisor, a scientist must be familiar with the administrative procedures of governments, with the rules of international diplomacy, and with the political obstacles that may have to be overcome if satisfactory solutions or agreements are to be reached. In other words, the scientific culture needs to be adapted to the cultures of professional diplomats and politicians.

On the other hand, he also discovered that his previous training as a scientist was very useful in his diplomatic work. When a particular issue arises, a professional diplomat is inclined to concentrate on finding the best way to represent the position of his or her government on that issue. He, or she, is trained to consider his or her own opinion to be of lesser relevance, and therefore expects and asks for explicit instructions for the negotiation.

The average scientist, however, is trained to find out for him- or herself about the merits of a case, and to form an opinion of his own. This enables the scientist to give independent advice to his or her superiors within the context of the governmental instructions. Even though the political instructions will not always allow him to express his opinion freely at the negotiation table, the fact that he has formed his view independently gives him greater confidence and a self-assured manner. His opponents will perhaps be more inclined to take him seriously, and his own government may be willing to take his suggestions into account. Even in the case of clear instructions, there is often some flexibility in the position to be negotiated that an intelligent negotiator can use for the benefit of his or her own side. Moreover, the informality, the openness, and the friendly behavior that usually characterize discussions among scientists and to which scientists, unconsciously, are accustomed, are helpful in creating a climate in which even difficult problems can be discussed productively. Cultural differences that usually make mutual understanding difficult play a lesser role among scientists, who share the common culture of science, than among representatives of some other professional cultures. In addition, scientists share the "elite culture" of those who travel frequently, know several languages, and are used to the exchange of ideas and arguments. Thus, their scientific background can assist scientists in obtaining satisfactory results in diplomatic negotiations. Scientists from opposing delegations in a negotiation are often in a position to build bridges to cross any rift caused by national differences.

It should, however, not be concealed that a certain understanding and broadmindedness on the part of the superiors of a scientist–negotiator is sometimes required to take full advantage of the assets reflected by the scientific culture. Because of their relative independence, scientists may be reluctant to follow rules that they do not understand or of which they do not approve. They may want at least to have an explanation why it might be considered reasonable to accept a particular rule. Rather than offering that explanation, or waiving the rule, politicians and professional diplomats sometimes prefer to look upon scientists as "eggheads" who lack political understanding and cannot be fully relied upon to follow instructions. According to this opinion, scientists may satisfactorily qualify as government representatives if their task is well defined and lies within their own field of competence; otherwise it may be better not to pay too much attention to them.

Fortunately, this type of skepticism regarding the usefulness of scientists in political negotiations of a scientific background is not universal. Particularly in the United States, it is quite common for scientists to switch back and forth between teaching and research, government service, and work for industry. This is looked upon with approval by the U.S. scientific community and the U.S. government as well. However, in Germany, for example, this kind of flexibility is still rare, but it does occur. The role of scientists, and therefore also the general significance of professional culture, is not consistent in all nations.

## Scientists as Advisors

In international negotiations, usually, there exist several possible solutions to a problem, each with its specific advantages, disadvantages, risks and costs. Which of these solutions seems to be preferable, depends on the interests of the parties negotiating with each other. An obvious question is: Are the benefits, risks and costs associated with any one solution distributed equally among the negotiating parties and/or the governments and nations they represent, or will one party bear most of the burdens while another one will enjoy most of the benefits? Decision-makers aware of their long-range responsibilities should also ask: Will only the present generation have the advantages from the chosen solution whereas future generations will have to pay the costs? Given the complexity of many problems on the agenda of today's international negotiations, a considerable number of preparatory meetings of scientific experts are often required before clear-cut alternatives can be presented to the political decision makers or their representatives for testing at the negotiation table.

Expressed in terms of process analysis, scientists have a particular role to perform in the early stages of an international negotiation. Whenever a technical solution that seems necessary from an overall and long-range point of view, for instance in the area of environmental protection, turns out to be unacceptable politically, it will be necessary to determine carefully the nature of the obstacles. Thus the impact of scientific culture on an unfolding negotiation is likely to be particularly significant in agenda-setting and issue clarification.

Scientists, or their institutions, have commented on all sorts of technical and scientific issues dealt with at the negotiation table, sometimes invited, sometimes at their own initiative. They have also played a significant role in interstate negotiations on national security at the heart of traditional diplomacy. Institution-building in the United States is a good example. The U.S. National Academy of Sciences (NAS) has set up standing committees on International Security and Arms Control (CISAC), and on Science, Engineering, and Public Policy (COSEPUP). The members of these committees are prominent, independent scientists working in fields other than those under revision by their committees. However, with the support of the NAS, these committees have established special working groups that conduct in-depth research into projects requiring government decisions on funding and/or political action, for example, concerning options for the disposal of plutonium from dismantled nuclear weapons. Such studies by NAS committees are generally recognized as being both technically competent and independent. CISAC has, for example, given itself three tasks:

• to inform members of the NAS on questions of international security and arms control;

• to create a cadre of independent experts on international security and arms control; and

• to promote international communication and to offer advice to the U.S. government on international security and arms control. This regime has meant the creation of an instrument for participation in national policy processes in the security domain, which has also carried the scientific professional culture into that context. In order to uphold political neutrality, a basic principle is: no agreements, no joint declarations, no publicity.

After the NAS had set up CISAC in 1980, the U.S.S.R. Academy of Sciences established a similar committee. One of its main tasks was to communicate with its American counterpart. The two committees (U.S. and Soviet) met regularly twice a year, alternately in the United States and the Soviet Union. In an informal atmosphere, free from polemics, the practical problems of security policy were discussed, and the results were reported back to the two governments. In parallel to the official arms control negotiations, these informal discussions of independent scientists contributed significantly toward a mutual understanding of the motivations, goals, perceptions, difficulties, and preoccupations of both sides. This is an interesting example of meetings of representatives of the same professional culture functioning as a bridge between two rivaling powers dealing with a sensitive issue.

In September 1986, the president of the NAS, Dr. Frank Press, suggested that European scientists, selected by European science institutions, should take part in discussions of this kind. After all, the security of the United States and the Soviet Union could not be separated from that of European countries on whose soil the nuclear weapons of the two blocs were deployed. Dr. Press invited the West European academies and scientific societies to consider forming a West European scientific committee on security issues to partner CISAC and its Soviet counterpart. A first exploratory meeting of CISAC members with European scientists interested in security questions had already taken place in Washington in June 1986.

The Accademia Nazionale dei Lincei (ANL) in Rome was the first European science organization to accept the American suggestion. A three-day workshop was organized which discussed the following topics:

- the treaty between the United States and the U.S.S.R. to eliminate intermediate-range and shorter-range nuclear missiles;

- the perspectives of drastic reductions in strategic arsenals;

- the reconversion of weapon-grade fissionable material to peaceful uses; and

- the future of the Strategic Defense Initiative (SDI)—points of view from Europe.

Similar workshops followed and the number of participants grew each time. For example, in 1990 scientists from seventeen countries attended, including from the Chinese Academy of Sciences. On this particular occasion the topics discussed concerned issues of international scientific and technological cooperation, particularly in environmental protection, energy use, climate research, conversion of armaments production, verification of technical disarmament measures, the design of new security concepts, and in the relations between industrial and developing countries. In honor of the late Professor Edoardo Amaldi, former president of the Accademia Nazionale dei Lincei, these conferences were to be known as the International Amaldi Conferences of Academies of Sciences and of National Scientific Societies, which ever since have held yearly sessions in different countries.

A constant topic of discussion at Amaldi Conferences in recent years has been how scientific institutions can take a more active role in working out options for dealing with urgent global problems, such as:

- pollution of soil, water, and air;
- destruction of the ozone layer;

- heating of the atmosphere;
- desertification;
- the alarmingly rapid disappearance of animal and plant species;
- the human population explosion;
- food and energy shortages;
- mass migration;
- nationalism, racism, and ethnic "cleansing;"
- psychological, social, and economic instabilities;
- civil wars and weapons trade; and
- the threat of nuclear proliferation and of the misuse of nuclear materials.

These problems are interconnected. Attempts to solve one of them separately can, often adversely, affect the solution of the other problems. What is needed, therefore, is a truly interdisciplinary and international approach upon which international negotiations can be based. The national academies of sciences and the national scientific societies have, in principle, the tools for such an approach. They are interdisciplinary and they have close relations to counterparts all over the world. Their members include the best experts in all fields of the natural and the social sciences—and in many cases also in the humanities. Compared to politicians, these experts are more independent, having tenure and not being dependent on re-election. They can therefore afford to concentrate on long-term goals rather than on short-term measures that may be popular at the moment but detrimental in the long run. In other words, scientific culture, conceived of as a widely shared outlook and frame of reference, is playing an important role in paving the way for integrative negotiation approaches and forward-looking outcomes.

The realm of international cooperation of scientific institutions comprises several associations, projects, and series of conferences that promote the role of scientific culture in the search for solutions to global problems, which today threaten the peaceful development of humankind on this planet. Probably the most universal of these associations are The International Council of Scientific Unions (ICSU) and The Inter-Academy Panel on International Issues (IAP). ICSU's activities include studies, projects, experiments, and panels on ecosystem processes and biodiversity, on the relations between health and environmental pollution, on climate change, on world ocean circulation, on the global cycles of energy and water, on stratospheric processes, on global change and terrestrial ecosystems, on biodiversity and sustainable use of the Earth's biotic resources, to mention but a few.[1] In 1999, ICSU convened in Budapest, jointly with UNESCO, a World Conference on "Science for the Twenty-First Century: a New Commitment" which adopted a "Declaration on Science and the Use of Scientific Knowledge" with special reference to scientific knowledge for progress, to science for peace and development, and to science in and for society.

IAP, in close cooperation with ICSU, has similar objectives. IAP organized a Population Summit in New Delhi in 1993 and it contributed to the HABITAT II Conference of the United Nations held in Istanbul in June 1996. In May of 2000 IAP sponsored a "Year 2000 Conference of Academies" in Tokyo which addressed opportunities and challenges for a worldwide transition to demographic, economic, and environmental sustainability in the 21st century. The use of electronic communications for the exchange of ideas and for online conferences is envisaged.[2] ICSU and IAP are not the only international and interdisciplinary efforts by the scientific community to study global problems and to inform governments and the public of them. The joint research programs of the European Science Foundation (ESF), among them one on Environment and Health, also play an important role.

In this context, the Pugwash Conferences on Science and World Affairs should also be mentioned (Gottstein 1986). The conferences deal with the threats to the survival of civilized humanity and to the health of future generations resulting from the existence of weapons of mass destruction, and with the options for eliminating these dangers. The purpose of these conferences is to bring together, from around the world, influential scholars and public figures concerned with reducing the danger of armed conflict and seeking cooperative solutions for global problems. Meeting in private as individuals rather than as representatives of governments and institutions, Pugwash participants exchange views and explore alternative approaches to arms control and tension reduction with a combination of candor, continuity, and flexibility seldom attained in official East-West and North-South discussions and negotiations. Yet, because of the stature of many of the Pugwash participants in their own countries (as, for example, science and arms-control advisors to governments, key figures in academies of science and universities, and former and future holders of high government office), insights from Pugwash discussions tend to penetrate quickly to the appropriate levels of official policymaking.[3]

By late 2002, there have been over 275 Pugwash conferences, symposia, and workshops, with a total attendance of over 10,000. There are now over 3500 individuals who have attended a Pugwash meeting (Source: Pugwash Website: www.pugwash.org). The first part of Pugwash's history coincided with some of the darkest years of the Cold War, marked by the Berlin Crisis, the Cuban Missile Crisis, the invasion of Czechoslovakia, and the Vietnam War. In this period of strained governmental relations, the fora and lines of communication provided by Pugwash played useful background roles in laying the groundwork for the Partial Test Ban Treaty of 1963, the Non-Proliferation Treaty of 1968, the Anti-Ballistic Missile Treaty of 1972, the Biological Weapons Convention of 1972, and the Chemical Weapons Convention of 1993. Subsequent trends of generally improving East-West relations and the emergence of a much wider array of unofficial channels of communication have provided alternate pathways to similar ends, and have thereby somewhat reduced Pugwash's visibility. But Pugwash meetings continue to play an important role in bringing together key analysts and policy advisors for sustained, in-depth discussions of the crucial arms-control issues of the day: European nuclear forces, chemical and biological weaponry, space weapons, conventional force reductions and restructuring, crisis control in the Third World, and terrorism. Pugwash has, moreover, extended its remit to include problems of development and the environment (Gottstein 1986).

Another example of a significant institution functioning in a bilateral context as an authoritative bearer of the scientific culture was the German–American Academic Council Foundation (GAAC). Founded following a joint announcement by the then German chancellor, Helmut Kohl, and the then U.S. president, Bill Clinton, in 1993, the foundation was dedicated to strengthening German–American cooperation in all fields of science and the humanities, particularly by bringing together and utilizing the experience, expertise, and commitment of its members. The foundation provided a forum for transatlantic dialogue, conducted policy studies of mutual interest to decision makers in both countries, and encouraged the development of collaborative networks, especially of young scientists and scholars.[4] Working groups have completed studies on the elimination of excess weapons plutonium, and on German and American migration and refugee policies. Symposia were held on academic and policy issues confronting Germany and the United States, on the future of humanities and social sciences in Central East European countries with consideration of experiences from the German unification process, on new horizons in research and higher education, and on scientific research in universities, academies, and specialized institutes in Central and Eastern Europe, Germany, and the United States. For German scientists, GAAC offered an opportunity to take part in interdisciplinary investigations of pressing problems. Unfortunately, it fell victim, in 1999, to economizing by the Schroeder government and had to terminate its very useful activities.

# Conclusions

This book proposes that the role of culture in international negotiations is not just restricted to the impact of national cultures. Professional cultures are also important conditioning factors influencing the initiation, development, and outcome of international negotiations. The possibility has to be considered that national and professional cultures may even compete with each other. The growing specialization in most areas of human action may enhance the importance of professional cultures in international cooperation and negotiation. Winfried Lang mentions the "lawyer–diplomat" to illustrate a representative of a highly professional culture. The "scientist–diplomat" is another example that seems to be becoming increasingly important in the current world situation.

In his part of the introduction to this book, Guy Olivier Faure states: "In an international negotiation, where national culture is the discriminating variable, a common professional culture can be viewed as a first step already made in the learning process that will enable the parties involved in a negotiation to reduce their divergence." This is certainly true for the scientific culture. It is not so much a particular vocabulary or a particular pattern of behavior, nor even similar skills or philosophical beliefs characterizing the international scientific culture that represent the key attributes of the scientific culture. The crucial factors are rather a profound respect for facts, an urge to find out about the driving forces behind the phenomenological world, a skepticism about unproven assertions, a contempt for lies, and a desire to make progress and see results, which are elements of scientific culture and which sometimes enable scientists to contribute usefully to negotiations and, in special cases, even to break deadlocks.

**Acknowledgements**

This chapter is an edited and updated version of Chapter 21 in: Klaus Gottstein, Catastrophes and Conflicts: Scientific Approaches to their Control, Ashgate Publishing Limited, Aldershot 1999. I thank Ashgate for permission to reproduce it here in adapted form. I am much indebted to Victor Kremenyuk who suggested that I should write this chapter. I am grateful to Marie Tweed for critically reading the text and making some useful suggestions.

# Notes

1. For a survey of the activities of ICSU, see ICSU (1996).
2. IAP Newsletter No 1, March 1996.
3. Pugwash conferences on science and world affairs, (A brief description), 1995, text issued by the Pugwash Central Office, London.
4. From the impressum of GAAC (1995).